\documentstyle[art12]{article}
\newcommand{\be}{\begin{eqnarray}}
\newcommand{\ee}{\end{eqnarray}}
\begin{document}
\thispagestyle{empty}
\large
\begin{flushright}UAHEP942\\
\end{flushright}
\begin{center}
     Running of the Top Yukawa with and without Light Gluinos$^*$\\
                             L. Clavelli\\
\small
                Department of Physics and Astronomy\\
                       University of Alabama\\
                     Tuscaloosa, Alabama 35487\\
\large
\medskip
\end{center}
\begin {abstract}
We investigate correlations among various parameters in the
solution space of minimal supersymmetric grand unification.
In particular the extent to which the top quark Yukawa
coupling exhibits fixed point behavior is discussed and we
compare various analytic approximations to its value at the
top mass with its exact value in numerical solutions.
\end{abstract}
\bigskip
\normalsize
\setcounter{page}{1}
\par
     One of the successes of the idea of supersymmetric grand
     unification is
the prediction of the $b$ quark to $\tau$ lepton mass ratio and the
accompanying
prediction of a top quark mass significantly above that of the $Z$.
These
predictions are based on the solution of a set of coupled renormalization
group differential equations involving the gauge and Yukawa couplings.
At least in the case of small $\tan(\beta), (<5)$, the solution space of
minimal SUSY
unification is a ten dimensional space defined by the values of the
following ten parameters.\\

\indent\indent     1)  A unification scale $M_X$\\
\indent\indent     2)  A unified gauge coupling $\alpha_0(M_X)$\\
\indent\indent     3)  A top Yukawa at $M_X$, $\alpha_t(M_X)$\\

\medskip
\footnotesize
\noindent
$^*$ Talk presented at the Workshop on Yukawa Couplings and the
Origin of Mass, Gainesville, Florida, Feb. 11-13, 1994\\
\normalsize
\vfill
\indent\indent     4)  A Susy scale $M_S$\\
\indent\indent     5)  A ratio $\tan(\beta)$ of the Higgs vacuum
expectation values\\
\indent\indent     6)  The weak angle $\sin^2(\theta_W)$\\
\indent\indent     7)  The fine structure constant $\alpha(M_Z)$\\
\indent\indent     8)  The strong coupling constant $\alpha_3(M_Z)$\\
\indent\indent     9)  The value of the top quark mass $M_t$\\
\indent\indent     10) The value of the $b/\tau$ mass ratio at the
$b$ quark scale\\
\par
     A "solution" is defined as a set of values for these ten
     parameters which
is consistent with the renormalization group running and with
the experimental
constraints:\\
\be
                    \alpha(M_Z) = 127.9\pm 0.2         \label{eq:az}
\ee
\be
                 \sin^2(\theta_W(M_Z)) = .2328\pm.0007\label{eq:sstw}
\ee
\be
                     m_b/m_\tau  = 2.39\pm0.10         \label{eq:b2tau}
\ee
The latter corresponds to a physical $b$ quark mass of $4.95\pm.15 GeV$.
There
are some who feel that the uncertainty in this quantity is much
smaller than
taken here but out of respect for the complications of confinement
we content
ourselves with this $4.5\%$ uncertainty.  The uncertainties on the
other two
quantities are below $1\%$ and once the top quark mass is known the
current
data
will specify them to about $0.15\%$ due to the correlation
\be
\sin^2(\theta_W) =.2324-.002( M_t^2 /(138 GeV)^2 - 1)\pm
.0003\label{eq:sstt}
\ee
The hope is that the increasing precision with which these numbers
are known
will shed light on the physics at the scales $M_S$  and $M_X$  and
will enable
predictions to be made for $m_t$  and $\alpha_3(M_Z)$.
\par
     In addition to the experimental constraints
eqs. \ref{eq:az},\ref{eq:b2tau},\ref{eq:sstt}  we assume that
$\alpha_t(Q) < 1$  for all $Q$ ("perturbativity") and that
\be
                            100 GeV < M_S  < M_{S,max}  \label{eq:ms}
\ee
If the Susy scale were below $100 GeV$, with the expected
degeneracy splittings
among the Susy particles, we would have expected unacceptably large
contributions to the $Z$ width due to Susy decay modes.  If $M_S$
is too large
the theoretical benefits of Susy in explaining the stability of a
low scale
for electroweak symmetry breaking is lost.  $M_{S,max}$ is variously
taken to be $1
TeV$ or $10 TeV$.  Ideally this argument would prefer $M_S$  below
one $TeV$
since the
electroweak scale is in the hundred $GeV$ range.  In addition, a Susy
solution
of the dark matter problem would require the mass of the LSP to
be no higher
than $200 GeV$ again suggesting an average Susy mass below one $TeV$.
Nevertheless we will, for the sake of conservatism, take $M_{S,max}=
10 TeV$.
\par
In the light gluino scenario, $\tan(\beta)$ is restricted to be between
     $1$ and
$2.3$\cite{CCY}    and, for simplicity, we limit our investigation
in the heavy gluino
case, to the range $1<\tan(\beta)<5$ so that we can neglect the
effect of the $b$
Yukawa on the running of the couplings.  This range is also preferred by
proton decay.  Radiative electroweak breaking would predict a value of
$\tan(\beta)$
very close to $1.8$ in the light gluino case.
\par
     In the simplest version of SUSY unification one assumes that
     all the GUT
scale particles are degenerate at $M_S$  and all the SUSY partners of the
standard
model particles are degenerate at $M_S$.  In the light gluino variant one
assumes
that the partners of the squarks and sleptons are at $M_S$  together
with the
heavy Higgs, while the photino and gluino are in the low energy
region below
$M_Z$  and the other neutralinos and charginos are at the scale of $M_Z$
as suggested
by the $M_{1/2}=0$ model.  This is consistent with the current
experimental gluino
searches which leave open (at least) the three windows shown in $fig.1$.
$Fig. 1$
updates the chart published by the $UA1$ \cite{UA1} group in 1987
to include the $LEP$
results which are probably the most model independent constraints
together
with the results of the $HELIOS$ \cite{Helios} collaboration which
searched for weakly
interacting neutral particles.
\par
     In higher level variations, the GUT scale spectrum is assumed to be
non-degenerate and possibly richer than in the minimal model and/or
the Susy
scale $M_S$ is split into different masses for the various particles.
In the
latter schemes for each non-degenerate spectrum of Susy particles
there is an effective degenerate scale $M_S$
which leads to the same unification solution
apart from small two loop effects.\cite{Langacker}    Each solution
in the degenerate case
corresponds to a family of solutions with different splittings
among squarks
and sleptons and, as long as $M_S$  is above $M_Z$ , no solutions
for the ten
parameters above are lost by assuming degeneracy.
\par
The two-loop differential equations to be solved are
summarized
in the papers of $refs.$\cite{Ramond,Barger}.
We follow a "top-down" approach where one begins by choosing random
values for the first 5 quantities in the list above, extrapolating
to low
energies and discarding the choices which are inconsistent with the
constraints of $eqs. $\ref{eq:az},\ref{eq:b2tau},\ref{eq:sstt},
\ref{eq:ms}.
The surviving solutions are stored in a data set
that can be queried for correlations among the various parameters.  The
solution space forms a small connected region in the ten dimensional
space.
Careful checking is required around the borders of this region to
insure that
these are in fact the borders and that no solutions exist outside
this region.
In $fig. 2a$ for example we show the solution space of the minimal
Susy model
projected onto the $M_S -M_X$  plane in a random non-overlapping
sample from $813$
solutions while $fig. 2b$ shows the same projection for a non-overlapping
subset of $1200$ solutions in the light gluino scenario.  The shape
coding
labels the $\alpha_3(M_Z)$ value of each solution.  In the heavy gluino
case solutions are found only for $0.111<\alpha_3(M_Z)<.134$
while in the light gluino case the range
is $0.122<\alpha_3(M_Z)<.133$.  These ranges are divided into quadrants
indicated in
the solutions of $fig. 2$ by rectangles, triangles, ovals, and
diamonds for
$\alpha_3(M_Z)$
in the lowest to the highest quadrant respectively.  One sees that
with the
light gluino option, The GUT scale is restricted to values
comfortably above
$10^{16} GeV$ while this is not the case in the standard Susy
picture.  In
Susy
unification, proton decay via lepto-quark gauge bosons of mass
as low as
$10^{15} GeV$ is not in contradiction with current limits. However,
proton decay via the
super-heavy scalars requires these particles to have masses in excess of
$10^{16} GeV$.  Since the GUT scale, above which the theory is
grand-unified,
is the
maximum mass of the GUT scale particles, solutions with $M_X$
below $10^{16} GeV$  are
probably not acceptable.  One could therefore add a sixth experimental
constraint
\be
                      M_X  > 10^{16}   GeV    \label{eq:mx}
\ee
which would cut the solution spaces of $figs. 2a,b$
at the corresponding limit.
The solution sets of $fig. 2a,b$ were generated in
the course of work reported in $ref.$\cite{CCMNW}.
One sees in $figs. 2a,b$ the effect reported there that the $\alpha_3$
values in
minimal Susy unification are significantly more constrained in the light
gluino scenario than in the usual picture.  This result, however, is
critically dependent on the super-gravity inspired prediction that
the charginos and neutralinos are relatively light (below the $Z$) when
the gluino is light.  In these data sets, the
constraint
of $eq. $\ref{eq:sstw}  was not enforced, although
that of $eq.$\ref{eq:sstt} was, so that a Susy prediction for
$\sin^2(\theta_W)$ could be made.  This prediction can be read from
$figs. 3a,b$ which show
the solutions in the $\sin^2(\theta_W)-\alpha_3(M_Z)$ plane.  One sees
that in both heavy and
light gluino cases, the predicted values of $\sin^2(\theta_W)$
lie between $0.230$ and
$0.2325$.  Thus the minimal model predicts the weak angle with a
$1\%$ accuracy and
agrees with experiment.  The shape coding indicates the quadrant
values of $M_S$
in the range from $100 GeV$ to $10 TeV$ (rectangle, triangle,
oval, diamond from
lowest to highest quadrant).  In figures $4a,b$ we show the heavy
and light
gluino solutions respectively in the $\tan(\beta)-M_t$  plane with the
values of $M_S$
indicated by the shape coding.  One sees that in both cases the
top quark mass
is bounded below by about $143 GeV$ and that in the light gluino case
the band
of solutions is appreciably broader.  It is interesting to note that the
preliminary evidence from Fermilab for $M_t \simeq 174 GeV$ suggests
a $\tan(\beta)\simeq 1.8$ as
required in the light gluino scenario with radiative electroweak
breaking.\cite{Nano}
In the current work, however, and that of \cite{CCMNW}, radiative
breaking is not assumed.
\par
     Much has been written \cite{Barger,Ibanez,Carena}
about the quasi-fixed point behavior of the top
Yukawa.  This can be roughly defined by the statement that the value of
the top Yukawa at the top mass given by
\be
\alpha_t(M_t) = M_t^2\cdot (173 GeV \sin(\beta)(1+4\alpha_3(M_t)/(3\pi)
                    +11(\alpha_3(M_Z)/\pi)^2))^{-2}  \label{eq:at}
\ee
is dependent only on the gauge couplings at the top scale and is
independent
of the GUT scale parameters at least for some range of those
parameters that
is consistent with the Susy unification solutions.  Our present
purpose is to
clarify and quantify this statement.
\par
     Neglecting the effect of the $b$ and $\tau$ Yukawas on the
     running of
$\alpha_t$  and
two loop contributions, the top Yukawa in the Susy region satisfies
\be
 2\pi d\alpha_t/dt = \alpha_t (6\alpha_t  - c_{t,i}\alpha_i)
 \label{eq:datdt}
\ee
where the $\alpha_i$ are the three gauge couplings, $t=ln(Q)$, and
\be
        c_{t,i} = (13/15 , 3, 16/3)           \label{eq:cti}
\ee
for $i=1,2,3$ in that order.  If $\alpha_t$  is initially
(at the GUT scale) higher than  $c_{t,i}\alpha_i/6$
the naive prediction is that it will fall with decreasing $Q$
until the right hand side of $eq. $\ref{eq:datdt} vanishes.
If $\alpha_t$  is below
$c_{t,i}\alpha_i/6$
initially, it will rise toward that value as $Q$ decreases.
At the most naive
level, one might expect from $eq. $\ref{eq:datdt} that $\alpha_t(Q)$
will approach
the fixed point
expression
\be
    \alpha_{t;f}^{(1)}(Q) = c_{t,i}\alpha_i/6 \label{eq:atf1}
\ee
{}From our unification solution set it is a simple matter to calculate
for each
solution the top Yukawa at $M_t$  from $eq. $\ref{eq:at}.  The gauge
couplings at $M_t$
can be related to the $\alpha_i(M_Z)$ from the renormalization group
expressions.  We can
therefore check how well $\alpha_t$ approaches $eq. $\ref{eq:atf1} at
$Q=M_t$ .
In figures $5a,b$, for
the heavy and light gluino scenarios respectively, we show the
correlation
between $\alpha_t(M_t)$ and $\alpha_{t;f}^{(1)}(M_t)$.  The correlation
is far from the close equality
one might have expected.  The shape coding, rectangles, triangles, ovals,
diamonds, indicates an $M_S$ value in the lowest to the highest quadrant
respectively with the total range being $100 GeV < M_S < 10 TeV$.
We will return
later to discuss the tail of solutions out to high
$\alpha_{t;f}^{(1)}(M_t)$
evident in figures
$5a,b$.
\par
Ibanez and Lopez\cite{Ibanez}    have shown that $eq. 8$ is satisfied
by a top Yukawa given by
\be
  \alpha_t(Q) = {- \alpha_t(M_X) \dot{F}(Q)
  \over{ 1+6\alpha_t(M_X)F(Q)/4\pi}}\label{eq:atfibanez}
\ee
where
\be
 F(Q) = - \int_{M_X}^Q {dQ'\over{Q'}}
 \exp(-\int_{M_X}^{Q'}
 {dQ''\over{4\pi Q''}}c_{t,i}\alpha_i(Q''))      \label{eq:fq}
\ee
and
\be
   \dot{F}(Q)\equiv Q {d \over{dQ}} F(Q)         \label{eq:fqdot}
\ee
$Eq. 11$ is of course not an analytic solution of $eq. $\ref{eq:datdt}
since the $Q'$ integral in
$eq. 12$ is not analytically soluble.  However it does illustrate the
quasi-fixed-point behavior in that $\alpha_t(Q)$ becomes independent
of $\alpha_t(M_X)$ if
$F(Q)$ is sufficiently large as would happen, for example, for large
enough $M_X$.
{}From $eq. $\ref{eq:atfibanez} one can write a second expression for
the fixed point by dropping the $1$ in the denominator
of $eq. $\ref{eq:atfibanez}.
\be
\alpha_{t;f}^{(2)}={-2\pi\over{3}}Q{d\over{dQ}}\ln F(Q)
\label{eq:atf2}
\ee
Because we don't have here an analytic solution for $F(Q)$, it is
difficult to predict the extent of the independence of $\alpha_t(M_t)$
on $M_X$ and $\alpha_t(M_X)$.  $Eq. $\ref{eq:atf2} does
indicate dependence of $\alpha_t(M_t)$ on $M_X$ and indirectly
on $\alpha_t(M_X)$ through the two
loop effects on the running of the gauge couplings.
The empirical dependence of $\alpha_t(M_t)$ on $M_X$ is substantial as
can be shown by projecting the solution space onto the
$\alpha_t(M_t)- M_X$
plane.  The Ibanez-Lopez
expression for the quasi-fixed-point does not yield analytically
the dependence of $\alpha_t(M_t)$ on the ten parameters of the
unification
solution.  We seek preferably a quasi-fixed-point expression
analogous to that of $eq. $\ref{eq:atf1}.
\par
     One of the problems with both of the above treatments is that
     the running
of the top Yukawa changes dramatically below the Susy scale.  Then
instead of
$eqs. $\ref{eq:datdt} and \ref{eq:cti} we have
\be
 2\pi {d\alpha_t^{sm}\over{dt}} =
 \alpha_t^{sm} ({9\over2}\alpha_t^{sm} - c_{t,i}^{sm} \alpha_i)
 \label{eq:datdtsm}
\ee
where below $M_S$  one defines the effective top Yukawa
\be
   \alpha_t^{sm} = \alpha_t \sin^2\beta   \label{eq:atsm}
\ee
and
\be
  c_{t,i}^{sm} = \big(17/20, 9/4, 8\big) \label{eq:ctism}
\ee
If $M_S$  is above $M_t$  one extrapolates from $M_S$  to $M_t$
using $eqs. $\ref{eq:datdtsm} and \ref{eq:ctism} and
then redefines $\alpha_t(M_t)$ through $eq. $\ref{eq:atsm} before
substituting in $eq. $\ref{eq:at}.  This
rescaling has only a higher order effect on the final answer so we use a
rescaling by the fixed value $\sin \beta=.7641$ in both the numerical
running and the
analytic expressions below.
\par
     Thus, if the top Yukawa is following a quasi-fixed-point given
approximately by $eq. $\ref{eq:atf1} down to $M_S$, below the Susy scale
we would expect it to be drawn toward a naive fixed point corresponding
to
\be
\alpha_{t;f}^{(1)sm}(Q) = {2\over{9}} c_{t,i}^{sm}\alpha_i(Q)
\label{eq:atfsm}
\ee
\par
   The corresponding $\alpha_{t;f}$ is more than twice the fixed point of
the Susy
regime suggested by $eq. $\ref{eq:atf1}.  This effect could partially
explain the large deviations of $\alpha_t(M_t)$ from
$\alpha_{t;f}^{(1)}(M_t)$ evident in figures $5a,b$.  However, it
is clear
that even with $M_S$  as large as $10 TeV$, the top Yukawa does not have
time to reach a standard model quasi-fixed-point.  From figures $5a,b$
one sees a tendency for the top Yukawa to rise with increasing $M_S$ as
would be expected
from this effect but it never achieves the doubling expected naively from
$eq. $\ref{eq:atfsm}.  Thus, if $M_S$ is in the $1$ to $10 TeV$ region,
the top quark Yukawa is unlikely to have reached a limiting behavior.
The tail of events at large $\alpha_{t;f}^{(1)}$ in figure $5$
represents the events in which $M_S <M_t$  so the Susy quasi-fixed-point
should
be most accurately attained.  Paradoxically, it is here that the
discrepancy
between $\alpha_t(M_t)$ and $\alpha_{t;f}^{(1)}$ is largest.  Clearly a
more accurate representation of the top Yukawa behavior is required.
\par
     In the SUSY regime the gauge couplings change according to the law
\be
     2\pi {d\alpha_i\over{dt}} = \alpha_i^2 b_i\label{eq:daidt}
\ee
with
\be
        b_i = (33/5, 1, -3)                      \label{eq:bi}
\ee
We can combine $eqs. $\ref{eq:datdt} and \ref{eq:daidt} with
arbitrary parameters $\lambda_i$
to write
\be
   2\pi {d(\alpha_t-\lambda_i \alpha_i) \over{dt}} = 6 \alpha_t^2
  - c_{t,i}\alpha_i\alpha_t - \lambda_i \alpha_i^2 b_i\label{eq:datmaidt}
\ee
where summation over repeated indices is intended and,
for the purpose of an analytic approximation, we have again neglected
two loop contributions to the running.  Ideally, one would seek
$\lambda_i$ such that when $\alpha_t$  reaches
\be
       \alpha_{t;f} = \lambda_i \alpha_i   \label{eq:atf}
\ee
the right hand side of $eq. $\ref{eq:datmaidt} would vanish.
Thus we would seek solutions to the equation
\be
 6(\lambda_i \alpha_i)^2 -
 c_{t,i}\alpha_i \lambda_j \alpha_j  - \lambda_i \alpha_i^2 b_i = 0
 \label{eq:lai}
\ee
However, no set of $\lambda_i$  exists that will
satisfy $eq. $\ref{eq:lai}
for arbitrary values of the $\alpha_i$.  We propose therefore to write
$eq. $\ref{eq:datmaidt} in the form
\be
   {\pi\over3} {d(\alpha_t - \lambda_i \alpha_i)\over{dt}} =
    (\alpha_t - c_{t,i}\alpha_i/12)^2 - \delta^2  \label{eq:datmaidt2}
\ee
where
\be
 \delta^2 = (c_{t,i}\alpha_i/12)^2 + \lambda_i b_i \alpha_i^2/6
 \label{eq:delta}
\ee
We will now choose
\be
        \lambda_i = c_{t,i}/12   \label{eq:lambdai}
\ee
and make the approximation of ignoring the $Q$ dependence of $\delta$.
Then $eq. $\ref{eq:datmaidt2}
can be integrated to write
\be
\alpha_t(Q) = \lambda_i \alpha_i(Q) + \delta {y_0+\delta +
(y_0-\delta) e^{6\delta\ln(Q/M_X)/\pi}
\over{y_0+\delta - (y_0-\delta) e^{6\delta\ln(Q/M_X)/\pi}}}
\label{eq:atf3}
\ee
where $y_0 \equiv \alpha_t(M_X)-\lambda_i \alpha_i(M_X)$ and $\delta$
is defined at $Q$.  This is an identity at $Q=M_X$
and gives an excellent approximation to $\alpha_t(Q)$ for all $Q>M_S$ .
In the limit $M_X/Q \rightarrow \infty$,
the exponentials in $eq. $\ref{eq:atf3} become negligible
and $\alpha_t(Q)$ is determined solely by the $\alpha_i(Q)$ becoming
independent of the GUT scale values.  In this
case one could talk about a quasi-fixed-point behavior.  However in
practice
$M_X$  is never large enough to justify dropping the exponentials.
Furthermore,
when $M_t <M_S$  we must face the complication of integrating $\alpha_t$
in the standard
model region.  In this case we use $eq. $\ref{eq:atf3} for
$\alpha_t(M_S)$
and write the approximate form using $eq. $\ref{eq:datdtsm}
\be
  \alpha_t(M_t) = \alpha_t(M_S)\bigg( 1 +{\ln{M_t/M_S}\over{2\pi}}\Big
  (0.7641 \cdot{9\over2} \alpha_t(M_S) - c_{t,i}^{sm}\alpha_i(M_t)
  \Big)\bigg)
  \label{eq:atfinal}
\ee
In $fig. 6a,b$ we show the correlation between the right hand side
of this
equation which we may call $\alpha_{t;f}^{(3)}$ and the exact numerically
calculated
left hand side.  The prediction is seen to hold within a few
percent over the
entire solution space.  Given $M_t$  and $M_S$ , the gauge couplings at
those scales
can be found to sufficient accuracy by the first order extrapolation from
their values at $M_Z$ .  If the exponentials in $eq. $\ref{eq:atf3} were
negligible,  $eqs. $\ref{eq:at} and \ref{eq:atfinal},  together with
knowledge of $M_t$ , $M_S$  and the gauge couplings at the $Z$ would
yield an accurate measure of $\tan(\beta)$ independent of the GUT scale
parameters.
However, only to a crude approximation $\simeq20\%$ can the
exponentials in $eq. $\ref{eq:atf3} be neglected.
\vskip 1cm
\par
     The author is indebted to Philip Coulter for discussions
     in the course
of this work.  The research reported here was supported in part by the
Department of Energy under Grant No. $DE-FG05-84ER40141$.

\vfill
\newpage
\centerline{\bf    FIGURE CAPTIONS }
\small
\smallskip
\noindent $Fig. 1.$\hspace{1mm}
  Low mass windows for the gluino mass as a function of the squark
         masses.  The hatched areas are disfavored by the indicated
         experiments. The dot-dashed curves represent the
         loci of expected
         gluino lifetimes $10^{-6}s, 10^{-8}s, 10^{-10}s,$
         and $10^{-12}s$ respectively
         from the highest to the lowest curve.\\

\noindent$Fig. 2.$\hspace{1mm}
  $(a)$ The correlation between the SUSY scale, $M_S$, and the GUT scale,
         $M_X$, in the heavy gluino case, $m_{\tilde g} = M_S$.  Each
         solution
         corresponds to
         an allowed point in the ten dimensional space discussed in the
         introduction.  For given $M_S$, no solutions exist outside of
         the broad
         band shown.  The width of the band corresponds to summing
         over all
         other eight parameters.  The $\alpha_3(M_Z)$ value for each
         solution is indicated by the shape coding (see text).\\
          $(b)$ The same correlation in the case of the light gluino
         ($m_{\tilde g} < M_S$).\\

\noindent$Fig. 3.$\hspace{1mm}
  $(a)$ The correlation between $\alpha_t(M_t)$ and $\sin^2(\theta_W)$
in the heavy gluino
         case.  Solutions in the 1st, 2nd, 3rd, and 4th quadrant of
         the $M_S$
         range, $100 GeV < M_S  < 10 TeV$, are printed as rectangles,
         triangles, ovals, and diamonds respectively.\\
          $(b)$ The same as $(a)$ but for the light gluino scenario.\\

\noindent$Fig. 4.$\hspace{1mm}
  $(a)$ The correlation between $\tan(\beta)$ and $M_t$  in the heavy
gluino case.
         The $M_S$  quadrants are indicated as in $Fig. 3.$
  $(b)$ The same as $(a)$ but for light gluinos.\\

\noindent$Fig. 5.$\hspace{1mm}
  The correlation between the "naive" top Yukawa fixed point
$\alpha_{t;f}^{(1)}(M_t)$
  and the actual top Yukawa $\alpha_t(M_t)$ found in the
  numerical solutions
   in the heavy gluino case $(a)$ and in the light gluino
   case $(b)$. $10\%$
   to $20\%$ departures are observed.  The quadrant values of $M_S$  are
         indicated by the shape coding as in $fig. 3$.\\

\noindent$Fig. 6.$\hspace{1mm}
  The correlation between the approximate analytic value,
$\alpha_{t;f}^{(3)}(M_t)$,
  of the top Yukawa and the exact numerical value of $\alpha_t(M_t)$
  in the
  heavy gluino solution space $(a)$ and the light gluino solution space
  $(b)$.  Agreement within $2\%$ is found.  The quadrant
  values of $M_S$  are
         indicated by the shape coding as in $fig. 3.$\\

\end{document}